# Narrowband IoT: An Appropriate Solution for Developing Countries

Sudhir K. Routray, Habib Mohammed Hussein
Department of Electrical and Computer Engineering
College of Electircal and Mechanical Engineering
Addis Ababa Science and Technology University, Addis Ababa, Ethiopia
Email: {sudhir.routray, habibmohamed2001}@aastu.edu.et

*Abstract*—Internet of things (IoT) is very much attractive for several sensor based applications. It provides large coverage of the services with small amount of resources. Its applications span from the ordinary scenarios such as sensing in the common digital ecosystem to the far more complicated processes of modern manufacturing, agriculture, security provisioning, location tracking and health care. Several types of IoTs have been proposed for the recent applications. Narrowband IoT (NBIoT) is one of the economical versions of the IoTs. It is a low power wide area network technology and thus suitable for resource limited scenarios. In the developing countries, the resources are scarce and economical solutions are always preferable. Therefore, NBIoT is an attractive solution for the developing countries. In this article, we present its features and functions which make it suitable for developing countries. We also provide several sector based analysis which are suitable for the NBIoT deployment.

## I. Introduction

The Internet of Things (IoT) is now an integral part of the modern digital ecosystem. It has several ramifications and can be applied to a number of sensor based application scenarios. It was initially started as a value added system for cellular communication networks. However, now they appear in several cellular and non-cellular forms including sensor networks, control networks and independent networks. IoTs can be of different types depending on their features. One of the standardized forms of IoT is narrowband IoT (NBIoT). As the name suggests, this IoT needs narrow band frequencies for its operations. Due to the narrow band requirement, it has several advantages over other forms of IoTs. Green and sustainable technologies are in high demand due to their energy efficiency and environment friendly features. These low energy technologies are very popular in different forms of IoTs and sensor based applications. These days, green IoTs are well researched areas in which optimization of resources takes the central position [1] - [5]. In the developing countries, they have even more important roles due to the shortage of resources [1]. In the recent years, several works have been done on different types of IoTs to make them more energy efficient. In [1], Green IoTs have been studies from several technological aspects. In [1], NBIoT has been proposed as the inherent green technology for several applications. In [2], energy efficient and sustainable initiatives for IoT based smart world have been addressed. It provides several recent progressive trends and technologies which are able to make the entire IoT networks energy efficient. It also gives insight for the segment-wise greening processes in the framework of existing information and communication technologies (ICTs) across different application domains. In [3], a detailed survey of the IoT marketplace of last few years has been presented. It explains the recent industrial demands and applications of IoT and its associated technologies in different processes. It also elaborates through justified logic that manufacturing and automation will have a large share of industrial IoT. In [3], a similar survey is provided which focuses on the technological prospective of IoT. It emphasizes that the wide deployments need to be energy efficient for the practical reasons. In [4], a comprehensive survey on the practical aspects of IoT has been provided. It first deals with the enabling technologies that make IoT a reality and then show the main protocols which are instrumental in the IoT operations. Then it provides a long list of applications in which IoT can play a significant role. The authors pointed that the overall success of IoT is very much dependent on the efficiency of the entire IoT networks, their components and end devices. In [5] and [6], the energy aspects of 4.5G and 5G are dealt with detailed information. Green initiatives of 5G are discusses in depth in [5].

In [7], the recent trends of consumer electronics driven data communication networks are provided. It shows the volume of the resources needed for the global communication networks. The data storage and communication infrastructure required for the current ICT sectors too have been presented in this article. In [8], sustainability aspects of communication networks have been presented. It shows the carbon emission contributions of global communication sector. Finally, it provides outlooks for overall sustainability. In [9], research and development initiatives of 5G and IoT are presented in which energy efficiency and overall network optimizations are analyzed. In [10], the green radio communication technologies and emerging methods are highlighted with proper background information. It also provides insights for emerging networking technologies for energy efficiency. In [11], 5G energy efficiency related aspects are presented which discusses the general system characteristics and their overall effect on the system. In [12] and [13], energy efficiency related aspects of the Internet and wireless Internet such as mobile Internet and IoT related aspects in specific have been discussed. It also covers the energy aspects related to IoT. In [14], green technologies for clouds are analyzed. Cloud energy efficiency is essential for green IoTs as clouds are going to be their integral parts in the long term. IoT data management needs clouds for the efficient ICT activities.

In this article, we provide the main initiatives being taken to make the IoTs efficient in resource utilization. We present the main features of NBIoT, its deployment issues and the commonly expected application domains. We show that narrowband IoT (NBIoT) is certainly the right choice for the developing countries now. We present its main motivating features, deployment options and applications in different sectors.

The reminder of this paper is organized in three sections. In Section II, we present the energy efficient and greener version of the current IoTs. Then we show that these features are found in NBIoT. We also presented its deployment options and applications in different sectors. In Section III, we present the overall initiatives being taken in the recent times to make NBIoT energy and resource efficient. We showed its suitability for the developing countries. In Section IV, we conclude with the main points of this article.

## II. GENERAL FEATURES OF NBIOT

NBIoT is the low energy and low bandwidth version of IoT which is designed for the massive machine-to-machine communications. As the name suggests, it uses narrow bands for its different functions and operations. It needs a bandwidth of just 180 kHz to 200 kHz for its designated processes. In LTE Release 13 and 14, 180 kHz has been proposed as the operating bandwidth for NBIoT. It is a low power wide area (LPWA) technology which can save a lot of power when compared with other forms of IoTs. It is good for large scale economical deployment of IoT for different applications. In true sense, it is leaner, thinner and greener than other IoTs proposed in the recent years. It can be deployed in both the cellular and non-cellular forms. However, cellular forms are popular as they can use the existing cellular architectures for its operations. In LTE Release 13 and 14, it has been standardized according to the compatible LTE provisions and also proposed for connected living environments [15].

### A. Basic LPWA Features of NBIoT

NBIoT is an energy efficient version of IoT. Main attraction of NBIoT is its LPWA nature. It can save a large amount of energy and bandwidth of a network. Its architecture and protocols have been standardized in the recent releases of LTE for different environments. Deployment bandwidths of NBIoT have been agreed by the standardization committee in Release 13. Depending on the situation, different spectral bands can be used by different operators. According to Release 13 of 3GPP LTE, the maximum usable bandwidth of an end device is 200 kHz. In fact, for the communication purpose 180 kHz is used [16] – [20]. At this bandwidth, the upper limits of the uplink and downlink data rates are set at 150 kbps. Half duplex mode has been recommended for NBIoT communications [17]. In Release 14, these specifications have been further enhanced for advanced applications. The power from the transmitter is kept quite small so that a single battery can supply power to the NBIoT devices or nodes for more than 10 years. In fact, in the recent NBIoT standards, two power levels have been specified: 20 dBm and 23 dBm [1]. The sensitivities of the NBIoT sensors are really good. They can receive the signal at a power level as low as 64 dBm. These are the exclusive features of NBIoT which is not found in other versions of IoTs proposed so far.

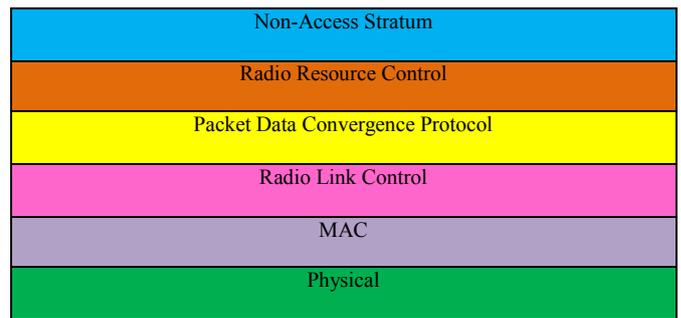

Fig. 1: OSI Model of NBIoT with six layers architecture.

### B. NBIoT Architecture

A systematic NBIoT architecture is required for its planning, dimensioning, cost estimation, design and final deployment. It does not have a legacy to follow as it is one of the earliest IoTs of its type. However, it is similar to the WSN and the WSNs are there for several years. The existing WSN architecture and topologies can be helpful in its further advancement. It is noteworthy that WSNs did not have a structured and well defined architecture like the cellular systems such as the LTE networks which will form the backbone of NBIoT. Therefore, an LTE cellular framework for NBIoT is the right choice at the moment [17]. The layered structure of NBIoT is helpful in its planning and deployment. In Release 13, several specifications for different layers have been mentioned. NBIoT can be separated into 6 layers as shown in Fig. 1. The physical layer is at the bottom and it is normally the air interface. Physical layer does the similar functions as other WSNs and some added functions as defined in Release 13. Above it is the medium access control (MAC) layer. This has the similar functions like the MAC layer of other networks. It incorporates the protocols for medium access and multiple access techniques. There is a radio link control layer in between the MAC and the upper layers. This layer makes the adaptation of the MAC layer information for radio links. Above it is the packet data convergence protocol layer which provides routing, traffic scheduling, networking and other related tasks. Then above it is the radio resource control layer which takes care of the radio resources of the packets in the channels. NBIoT uses user datagram protocol (UDP) and other cellular mechanism to carry this function [18]. UDP is effective in the wireless networks and thus suitable for NBIoT as well. The topmost layer is the Non-

Access Stratum (NAS) which establishes the communication between the user equipment (UE) and the main server of the NBIoT also known as NBIoT central node.

## C. Deployment Options for NBIoT

NBIoT deployment options have been agreed in Release 13. According to it, NBIoT can be deployed in three different forms. The first being the standalone deployment in which an independent microwave band is provided in the 700MHz or 800 MHz range for its deployment. The second one is the guard band deployment in which the spectrum for IoT is the unused GSM and LTE guard bands. These guard bands are normally not used for communication purposes in GSM and LTE cellular systems. With the arrival of NBIoT, these guard bands find a new application. The third one is the in-band deployment. In this case, some part of the GSM or LTE spectrum is allocated to the NBIoT. In Fig. 2, we show these 3 deployment scenarios. In practice, either of them can be used depending on the availability of the bands and its suitability. If required, even two or all three options can also be used in specific cases.

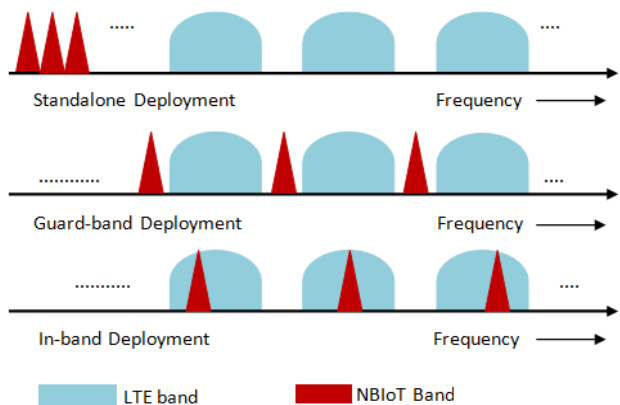

Fig. 2: NBIoT deployment options in different bands.

## D. Applications of NBIoT

NBIoT is attractive in the LPWA regime. It has tremendous potential to operate in the resource limited scenarios. Here, we shed the light on some of the typical sectors where it is currently considered for deployment. Several applications of NBIoT are possible and it can be deployed for any LPWA applications in which the six elements of IoT are available. Both local and nonlocal applications of NBIoT are equally possible. Emerging applications are found in several technology and social sectors with every passing month. There are several new sectors in which NBIoT has potential applications. Some of these applications are indoor and the others are outdoor. NBIoT can be instrumental in designing smart homes and smart cities [16]. Similarly, it can help in healthcare as well as the security monitoring of public places. In Fig. 3, we present some typical sectors in which NBIoT can have significant role. These sectors are: smart city management and monitoring; smart home applications; large scale manufacturing; pet tracking; kids' tracking; healthcare monitoring; safety and security related applications; smart agriculture and farming; energy and utility management; automobile and vehicular management; retail management;

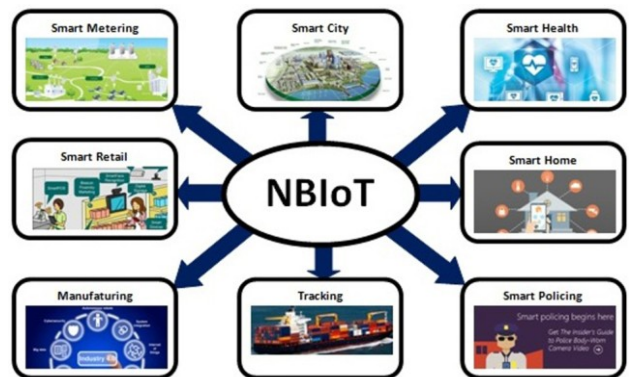

Fig. 3: Different applications of NBIoT. New applications emerge every month in different sectors.

policing and law enforcement assistance; detection of environmental degradation; waste management; and several other LPWA applications. In fact, for rural areas NBIoT is considered as the only choice due to its economical features.

## III. NBIOT FOR DEVELOPIONG COUNTRIES

In Section II, we have seen the attractive LPWA features of NBIoT. It is very much energy efficient and provides significantly large coverage which is not possible by other versions of IoTs. The LPWA features are suitable for resource limited scenarios. In developing countries resources are scarce and economical solutions are always an advantage. Furthermore, the legacy systems such as GSM are widely found in the developing countries. NBIoT is compatible with GSM as well as the recent versions of the LTE such as 4G and 4.5G. However, other forms of IoTs are compatible with only the recent versions of LTE. Therefore, NBIoT is a natural choice for the developing countries.

## A. NBIoT for Agriculture in Developing Countries

NBIoT can be deployed in the agricultural sector for several performance enhancement related aspects. In agriculture, several resources are used to increase the harvest. Many of those resources are underutilized due to the lack of facilities to monitor their utilities. For optimal use of the resources, NBIoT can be used which can sense the level of utilization and improve the utilization efficiency. For instance, water is an essential resource for agriculture. It is not available in sufficient amount in every part of the world. In the dry areas, water has to be utilized very efficiently. This utilization can be monitored in the agricultural farms through the water sensors deployed under the soil. Similarly, fertilizer concentration, pesticide intensity and humidity levels too can be monitored using different types of sensors as a part of NBIoT in agriculture. These resources can also be remotely controlled through the NBIoT. As the energy levels for these sensing operations are very low, the cost of this IoT based monitoring is very much economical for the farmers. On the other hand, the harvest can be maximized due to the efficient nutrition distribution to the plants. These initiatives are already being deployed in several countries and found to be very effective. Of course, it will be made even more efficient in the future.

*B. NBIoT for Manufacturing in Developing Countries*

Manufacturing is an important economic sector for developing countries. Recently, manufacturing companies in the developing countries are experiencing extreme competition, mainly due to the increasing pressures from technological changes and global business challenges. These pressures result in the globalization of manufacturing, characterized by faster transfers of materials, complex payment systems and the compression of products' life cycles, which drives the integration of technologies with increasingly sophisticated customers' needs [21]. To be successful companies try to anticipate future trends by developing ideas, products or services to rapidly and effectively meet future demands. In addition to that they are responding to their current customers' or organizational needs to sustain competitive advantage. Among all the sets of pressures of a technological nature, the advent of the Internet has deeply affected companies' approach to production and has strongly reshaped organizational and operational structures. However, the role of the Internet in manufacturing is still understudied as it is for the IoT phenomenon. Advanced manufacturing technologies strongly rely on various ICT technologies to achieve higher productivity, higher quality and lower production costs. Such an effect is especially focused on processes of manufacturing automation, and of information systems. Indeed, the advent of Internet-based technologies has led to the emergence of new manufacturing philosophies and new forms of organization, such as virtual organizations, remote manufacturing, computer-integrated manufacturing systems, and Internet-based manufacturing, i.e. wireless milling machines, coordinated measuring machines, networked sensor arrays and surveillance systems. For example, "design anywhere, manufacture anywhere" is a new approach to production which shares design and manufacturing data across multiple platforms and infrastructures [22]. Recent studies have confirmed such trends, indicating that the future of manufacturing firms will be mostly information-oriented and knowledge-driven, leading to a much more flexible and an abundance of automated operations systems.

In all these changing scenarios, overall efficiency and automation can be provided by NBIoT [23]. Thus, it is a hot choice for the industrial automation in the developing countries.

*C. NBIoT for Healthcare in Developing Countries*

Healthcare provisioning in any country is one of the basic services. It is also a large sector due to the very nature of the services provided by this system. Healthcare system is multi-tiered and need proper collaboration among each of the tiers. In several healthcare processes, NBIoT can play a role of game changer. For instance, in emergency proceedings it can help the patients to get appropriate treatments while coming to the hospital from their residence. Using the sensors, the vital information of the patients can be sent to the doctor who in return can provide the appropriate support through the sensors till the patient reaches the hospital. Similarly, the telemedicine applications too can be improved through the NBIoT based sensor assistance.

*D. NBIoT forResource Management in Developing Countries*

Basic resources such as water, electricity, gas and other essential commodities are not properly managed in the developing countries. That leads to the poor resource utilization efficiency and loss of revenue for the government and associated public sectors. NBIoT based solutions already exist to improve the resource utilization efficiency in these areas. Smart metering and efficient resource distribution can be a game changer in these sectors. NBIoT based automation in the metering and resource distribution is quite cost effective and efficient. It can provide a real-time resource monitoring framework to the resource distributors. This is how the operators can directly monitor the efficiency of resource utilization. Thus NBIoT is being deployed in several countries for the public resource management.

IV. CONCLUSIONS

Narrowband IoT is unique in the low power regime of the IoTs. It is also special in the wide coverage and low cost of its deployment. In this article, we have presented the main features of NBIoT, its architecture and an OSI model for its layer wise architecture which is needed in its design and implementation. NBIoT is very much suitable for the developing countries. Its low cost and wide coverage makes it the first choice for the common applications in the developing countries. NBIoT is very much effective in agriculture, manufacturing, healthcare, resource management and social applications such as policing. We presented the overall effectiveness of NBIoT in the developing countries. In the future there will be many new applications of NBIoT in the developing countries.